\documentclass[11pt,a4paper]{article} \usepackage{jheppub}

\usepackage{hyperref}
\usepackage{graphicx}
\usepackage{amsmath,amssymb,amsfonts,amsthm,latexsym,stmaryrd}

\def\be{\begin{equation}}
\def\ee{\end{equation}}
\def\ba{\begin{eqnarray}}
\def\ea{\end{eqnarray}}

\def\eps{\varepsilon}
\def\lp{\ell_\text{Pl}}

\def\de{\mathrm{d}}

\def\SU{\text{SU}}
\def\DSU{\text{DSU}}
\def\ISU{\text{ISU}}

\def\SL{\text{SL}}
\def\su{\mathfrak{su}}
\def\so{\mathfrak{so}}

\begin{document}

\title{Statistical Entropy of a BTZ Black Hole\\from Loop Quantum Gravity}

\author[a]{Ernesto Frodden,}
\affiliation[a]{Centre de Physique Th\'eorique\footnote{Unit\'e Mixte de Recherche (UMR 6207) du CNRS et des Universit\'es Aix-Marseille I, Aix-Marseille II, et du Sud Toulon-Var; laboratoire afili\'e \`a la FRUMAM (FR 2291).}, Campus de Luminy, 13288 Marseille, France.}

\author[b,c]{Marc Geiller,}
\affiliation[b]{Laboratoire APC -- Astroparticule et Cosmologie, Universit\'e Paris Diderot Paris 7, 75013 Paris, France.}
\affiliation[c]{Institute for Gravitation and the Cosmos \& Physics Department, Penn State, University Park, PA 16802, U.S.A.}

\author[d,b]{Karim Noui,}
\affiliation[d]{Laboratoire de Math\'ematique et Physique Th\'eorique\footnote{F\'ed\'eration Denis Poisson Orl\'eans-Tours, CNRS/UMR 6083.}, 37200 Tours, France.}
\affiliation[b]{Laboratoire APC -- Astroparticule et Cosmologie, Universit\'e Paris Diderot Paris 7, 75013 Paris, France.}

\author[a]{Alejandro Perez}
\affiliation[a]{Centre de Physique Th\'eorique\footnote{Unit\'e Mixte de Recherche (UMR 6207) du CNRS et des Universit\'es Aix-Marseille I, Aix-Marseille II, et du Sud Toulon-Var; laboratoire afili\'e \`a la FRUMAM (FR 2291).}, Campus de Luminy, 13288 Marseille, France.}

\emailAdd{efrodden@gmail.com}
\emailAdd{geillermarc@gmail.com}
\emailAdd{karim.noui@lmpt.univ-tours.fr}
\emailAdd{perez@cpt.univ-mrs.fr}

\abstract{
We compute the statistical entropy of a BTZ black hole in the context of three-dimensional Euclidean loop quantum gravity with a cosmological constant $\Lambda$. 
As in the four-dimensional case, a quantum state of the black hole is characterized by a spin network state. Now however, the underlying colored graph $\Gamma$ lives in a two-dimensional spacelike surface $\Sigma$, and some of its links cross the black hole horizon, which is viewed as a circular boundary of $\Sigma$. Each link $\ell$ crossing the horizon is colored by a spin $j_\ell$ (at the kinematical level), and the length $L$ of the horizon is given by the sum $L=\sum_\ell L_\ell$ of the fundamental length contributions $L_\ell$ carried by the spins $j_\ell$ of the links $\ell$. We propose an estimation for the number $N^\text{BTZ}_\Gamma(L,\Lambda)$ of the Euclidean BTZ black hole microstates (defined on a fixed graph $\Gamma$) based on an analytic continuation from the case $\Lambda>0$ to the case $\Lambda<0$. In our model, we show that $N^\text{BTZ}_\Gamma(L,\Lambda)$ reproduces the Bekenstein-Hawking entropy in the classical limit. This asymptotic behavior is independent of the choice of the graph $\Gamma$ provided that the condition $L=\sum_\ell L_\ell$ is satisfied, as it should be in three-dimensional quantum gravity.}

\maketitle

\section*{Introduction}

\noindent Twenty years ago, Ba\~nados, Teitelboim, and Zanelli discovered a black hole solution in three-dimensional space-time in the presence of a negative cosmological constant $\Lambda=-\ell_c^2$ \cite{BTZ}. This discovery came as an enormous surprise mainly because gravity in three dimensions does not have local degrees of freedom, whereas the black hole has thermodynamic properties analogous to those of higher dimensional black holes. In particular, a BTZ black hole radiates at a Hawking temperature $T_\text{H}$ and admits an entropy $S$, which means that it has a large number of microscopic degrees of freedom. The question of the origin of these degrees of freedom has lead to a huge literature, but the precise nature of these fundamental excitations of the black hole is still not totally understood (see \cite{Carlip} for a very interesting review). The key to the problem lie in the fact that the BTZ black hole admits a conformal field theory (CFT) description. More precisely, in the Lorentzian case, the BTZ black hole is locally isometric\footnote{In fact, it is globally defined as the coset of $\text{AdS}_3$ by a discrete subgroup of the $\text{AdS}_3$ isometry group $\SL(2,\mathbb{R})\times\SL(2,\mathbb{R})$ \cite{BHTZ}. Fundamentally, this result comes from the fact that solutions to three-dimensional gravity have constant curvature, this curvature being negative when $\Lambda<0$.} to the three-dimensional anti-de Sitter space-time $\text{AdS}_3$, and its asymptotic symmetries are generated by a pair of Virasoro algebras with a central charge $c=3\ell_c/G$, where $G$ is Newton's constant. This result, which was originally pointed out in \cite{Brown-Henneaux}, has been the very first example of the famous AdS/CFT correspondence. Starting from this observation, the knowledge of the central charge and the universality of the Cardy formula for computing the density of states are then the only necessary ingredients to recover the black hole entropy. Thus, it has been argued that the asymptotic CFT carries the microscopic degrees of freedom responsible for the statistical entropy of the BTZ black hole \cite{Strom,BSS}.

However, the precise nature of the microstates remains unclear despite all the work devoted to the statistical description of the BTZ black hole. Although it was shown that the asymptotic CFT is a Liouville theory \cite{CHD}, this theory does not have a priori enough degrees of freedom (normalizable quantum states) to explain the high value of the BTZ black hole entropy. Many arguments have been put forward in order to resolve this problem, and it has been suggested that the asymptotic Liouville theory is only an effective theory that cannot describe the fundamental excitations of the black hole \cite{Martinec}. From the technical point of view, the difficulty of finding the microstates is mainly due to the non-compactness of the symmetry group $\SL(2,\mathbb{R})\times\SL(2,\mathbb{R})$ of Lorentzian three-dimensional gravity with a negative cosmological constant. Indeed, this implies that the symmetry group of the asymptotic CFT associated to the BTZ black hole is also non-compact, which makes the quantization much more difficult than in the compact case. Attempts to properly quantize this theory in relation with the BTZ black hole have been developed, but the complete picture is still missing \cite{Chen}.

As the main difficulty is due to the non-compactness of the Liouville symmetry group, people have quickly tried to circumvent this problem by first looking at situations with more compact symmetries and then performing some analytic continuation. Even if this strategy is not fully satisfactory, it could give important indications about the quantization of the non-compact case. One way towards this simplification consists in changing the Lorentzian signature to the Euclidean one by a Wick rotation. Fortunately, there still exists a black hole solution in the Euclidean case when $\Lambda<0$, called the Euclidean BTZ black hole, and it admits thermodynamical properties as well. In this context, the black hole is locally isometric to the three-dimensional hyperbolic space $\mathbb{H}_3$. Globally, it is defined as the quotient of $\mathbb{H}_3$ by a discrete subgroup of its isometry group $\SL(2,\mathbb{C})$, and it has the topology of a solid torus \cite{BHTZ}. Furthermore, the Lorentz group $\SL(2,\mathbb{C})$ is, as expected, the symmetry group of Euclidean three-dimensional gravity with a negative cosmological constant $\Lambda$. Even if $\SL(2,\mathbb{C})$ is still non-compact, it contains $\SU(2)$ as a subgroup, and therefore can be thought of as being ``less'' non-compact than $\SL(2,\mathbb{R})\times\SL(2,\mathbb{R})$. In this way, the technical difficulties of the quantization are a bit tamed. Besides, it is possible to completely perform the canonical quantization of Chern-Simons theory with the Lorentz group \cite{BNR}. One can even go further into the ``compactification'' of the symmetry group, by simply trading the negative cosmological constant for a positive one. In this case, the symmetry group becomes $\SU(2)\times\SU(2)$. It is therefore totally compact, and the quantization of the Chern-Simons theory can be performed in various ways (the first and certainly the deepest one is due to Witten \cite{Witten_Jones}). Even if no black holes exist in this case, one can think of using results from $\SU(2)\times\SU(2)$ Chern-Simons theory in order to obtain results for Euclidean quantum gravity in the presence of a negative cosmological constant. This is exactly what has been done in \cite{Carlip-Euclidean} to obtain the number of black hole microstates by performing an analytic continuation of the $\SU(2)\times\SU(2)$ Chern-Simons partition function on a solid torus, to a negative value of the cosmological constant $\Lambda$. In this way, one recovers the black hole entropy $L/(4\lp)$.

The goal of this paper is to adapt this strategy in order to compute the BTZ black hole entropy in the framework of three-dimensional Euclidean loop quantum gravity. We proceed as follows. As in the four-dimensional case \cite{ENP}, the black hole horizon is described as a circular boundary in a spacelike surface $\Sigma$, and the kinematical states are spin networks associated to a two-dimensional graph $\Gamma\subset\Sigma$ with links crossing the horizon. For a given graph, we assume that $p$ (by analogy with the number of punctures in the four-dimensional case) links cross the horizon, and that these links are colored by unitary irreducible representations $j_\ell$ (for $\ell\in\llbracket1,p\rrbracket$) of the group $\SU(2)$ as it should be at the kinematical level (regardless of the value of the cosmological constant $\Lambda$). Each link carries a quantum of length $L_\ell=8\pi\lp\sqrt{j_\ell(j_\ell+1)}$, where $\lp=\hbar G$ is the Planck length, and the sum of the fundamental contributions gives the macroscopic horizon (one-dimensional) ``area'' $L=\sum_\ell L_\ell$. As kinematical states are necessarily $\SU(2)$ gauge-invariant, the number $N^\text{kin}_\Gamma(L)$ of kinematical microstates of the black hole horizon, which depends a priori on the colored graph $\Gamma$ and on the area $L$, is given by the number of $\SU(2)$ intertwiners between the representations $j_\ell$. To continue, we assume now that the cosmological constant is positive, i.e. $\Lambda=\ell_c^2$. We know, from the different quantization schemes of three-dimensional gravity \cite{Witten_Jones,Schom,Schom2}, that physical states (once the remaining constraints involving the curvature of the connection are imposed) are quantum spin networks colored with representations of $\text{U}_q(\su(2))$, where $q=\exp\big(i2\pi/(k+2)\big)$ is a root of unity, and the level $k$ is an integer given by $k=\ell_c/\lp$. Moreover, recent results \cite{Pranz1,Pranz2,Pranz3,Pranz4,Hanno} strongly indicate that the quantum group $\text{U}_q(\su(2))$ could emerge directly in the context on loop quantum gravity at the physical level, but a precise and complete proof of this fact is still missing. Therefore, at the physical level, since the representations coloring the graph $\Gamma$ are viewed as representations of $\text{U}_q(\su(2))$, they remain labelled by half-integers $j_\ell$ but are now bounded by $k/2$. Furthermore, the number $N_\Gamma(L,\Lambda)$ of physical microstates of the black hole horizon depends on $\Lambda$, and is now given by the number of $\text{U}_q(\su(2))$ intertwiners between the representations $j_\ell$. This number is well-known and has been studied quite a lot, in particular in the context of $\SU(2)$ black holes in loop quantum gravity \cite{entropyrevisited}. We propose an analytic continuation of $N_\Gamma(L,\Lambda)$ to a negative value of $\Lambda$, denote by $N^\text{BTZ}_\Gamma(L,\Lambda)$ the resulting number of microstates, and show that, in the classical limit where $\lp\rightarrow 0$ and $j_\ell\rightarrow\infty$ with $\lp j_\ell\rightarrow a_\ell$, the entropy behaves as 
\be
S^\text{BTZ}_\Gamma(L,\Lambda)=\log\left(N^\text{BTZ}_\Gamma(L,\Lambda)\right)\sim\frac{L}{4\lp}.
\ee
In this way, we recover the Bekenstein-Hawking formula for the BTZ black hole entropy. Furthermore, this result does not depend on the choice of $\Gamma$, as it is expected in three-dimensional gravity.

This paper is organized as follows. In the next section, we briefly review basic results about the BTZ black hole. In section \ref{sec:2}, we describe the (kinematical and physical) microstates of the BTZ black hole in the framework of three-dimensional loop quantum gravity. In section \ref{sec:3}, we estimate the number of microstates of the BTZ black hole, and show that in the classical limit it reproduces the Bekenstein-Hawking entropy formula. We conclude with a discussion of this result and of future investigations concerning for instance the fate of the logarithmic corrections in our approach.

\section{A brief overview of the BTZ black hole}
\label{sec:1}

\noindent Gravity in three space-time dimensions is a topological field theory. It has no propagating degrees of freedom, and locally the space-time has a constant curvature whose value depends on the cosmological constant $\Lambda$. Despite its apparent simplicity, three-dimensional gravity is far from being trivial, and the discovery \cite{BTZ} of the existence of a black hole solution when $\Lambda=-\ell_c^2<0$ is a nice illustration of its physical richness.

In Schwarzschild-like coordinates, the BTZ metric is given by
\be\label{BTZmetric}
\de s^2=N^2\de t^2-N^{-2}\de r^2-r^2(\de\phi+N^\phi\de t)^2,
\ee
where the (positive) lapse function $N(r)$ and the shift function $N^{\phi}(r)$ and defined by
\be
N^2=-8GM+\frac{r^2}{\ell_c^2}+\frac{16G^2J^2}{r^2},\qquad\qquad N^\phi=-\frac{4GJ}{r^2}.
\ee
Here $M$ and $J$ are respectively the mass and the angular momentum of the black hole, and they satisfy the inequality $|J|\leq M\ell_c$. The metric \eqref{BTZmetric} describes a space-time of constant negative curvature, and the BTZ black hole is therefore globally defined as a coset of the three-dimensional anti-de Sitter space-time $\text{AdS}_3$ by a discrete subgroup of its isometry group $\SL(2,\mathbb{R})\times\SL(2,\mathbb{R})$. Furthermore, it has an event horizon at $r_+$, where
\be\label{horizon}
r_\pm^2=4GM\ell_c^2\left(1\pm\sqrt{1-\frac{J^2}{M^2\ell_c^2}}\right). 
\ee
When the angular momentum $J$ is different from zero, the BTZ black hole also possesses an inner Cauchy horizon at $r_-$ that we will not consider in the following. The most important point for our problem is that this black hole admits thermodynamical properties similar to those of four-dimensional black holes. In particular, it has an entropy equal to one fourth of its area $L=2\pi r_+$ in Planck units, i.e.
\be\label{entropy}
S=\frac{L}{4\lp}.
\ee
These thermodynamical features can be derived following the same methods used in four space-time dimensions.

Furthermore, the Lorentzian BTZ black hole admits an Euclidean counterpart. This latter is a solution of Euclidean gravity with a negative cosmological constant. It defines an Euclidean space of constant negative curvature as well, and therefore is globally defined as the coset of the three-dimensional hyperbolic space $\mathbb{H}_3$ by a discrete subgroup of the Lorentz group $\SL(2,\mathbb{C})$. A precise study shows that it has the topology of a solid torus. The metric can be written in Schwarzschild-like coordinates (\ref{BTZmetric}) by performing the Wick rotation $t\rightarrow i\tau$, and also changing $J$ to $iJ$. The continuation of the angular momentum $J$ to purely imaginary values is necessary in order to keep the metric real. The Euclidean BTZ black hole still has an event horizon at $r_+$ (\ref{horizon}) provided that we replace $J$ by $iJ$, and possesses the same entropy $S$ (\ref{entropy}) as the Lorentzian black hole.

\section{The BTZ black hole in loop quantum gravity}
\label{sec:2}

\noindent From the point of view of loop quantum gravity, the BTZ black hole is characterized by its horizon at $r_+$, viewed as an isolated horizon \cite{ADW}. The situation is similar to the four-dimensional case. In this picture, space-time is locally isomorphic to $\Sigma\times I$ where $I$ is an interval of $\mathbb{R}$ and $\Sigma$ a surface with a circular boundary representing the black hole horizon. For the reasons that we have explained in the introduction, we start with Euclidean gravity in the presence of a positive cosmological constant. Then, we will perform an analytic continuation to the case of a negative cosmological constant, and discuss its physical interpretation.

The phase space of (first order) three-dimensional Euclidean gravity is parametrized by a triad field $e$ and an $\su(2)$ connection $A$, with Poisson bracket
\be\label{Poisson}
\{A_a^i(x),e_b^j(y)\}=8\pi G\eps_{ab}\eta^{ij}\delta^2(x-y).
\ee
Here $a,b,\dots$ denote the spatial indices, $i,j,\dots$ denote the $\su(2)$ internal indices, $x,y,\dots$ are spatial coordinates, $\eps_{ab}$ is the two-dimensional antisymmetric tensor, and $\eta$ is the flat metric. The canonical analysis of three-dimensional gravity implies that these variables are subject to the following first class constraints:
\begin{subequations}\label{cons}
\ba
G&\equiv&\partial_ae_b-\partial_be_a+A_a\times e_b-A_b\times e_a\approx0,\label{gauss cons}\\
H&\equiv &F_{ab}(A)-\Lambda e_a\times e_b\approx0,\label{curvature}
\ea
\end{subequations}
where $F(A)$ is the curvature of the connection $A$, and $(u\times v)^i=\eps^{ijk}u_jv_k$. The existence of these six first class constraints implies that there are no propagating local degrees of freedom.

\begin{figure}[h]
\begin{center}
\includegraphics[scale=0.5]{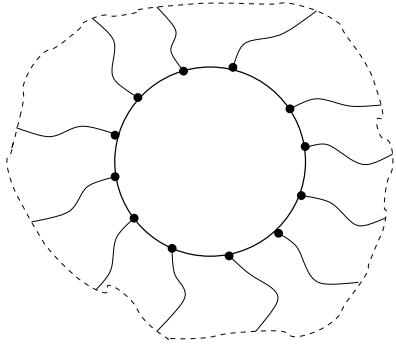}
\end{center}
\caption{The graph $\Gamma$ has $p$ links that cross the black hole horizon on the surface $\Sigma$. Outside of the horizon, the links are supposed to meet at only one point located at infinity.}
\label{graph}
\end{figure}

In loop quantum gravity, one first quantizes the Poisson bracket (\ref{Poisson}) by promoting the classical variables to quantum operators, and then the constraints (\ref{cons}) are implemented at the quantum level. Quantum states are taken to be functionals $\psi(A)$ of the connection $A$, and the tetrad field $e$ acts on them as a derivative operator. The loop assumption underlying the quantization scheme consists in considering only cylindrical function $\psi$ defined on graphs $\Gamma\subset\Sigma$. This assumption is justified in three-dimensional gravity because the theory is topological. Therefore, all the degrees of freedom are captured by a single graph provided it is sufficiently refined to resolve the surface topology. For this reason, we consider only one graph $\Gamma$ with $p$ links crossing the horizon, as illustrated in figure \ref{graph}, where $p$ is for the moment arbitrary. Furthermore, the topology of $\Sigma$ is rather simple since it is a plane $\mathbb{R}^2$ with a circular boundary. As a consequence, it is sufficient (in order to resolve the surface topology) to take a graph $\Gamma$ such that the links outside the horizon meet only at one point at infinity. This is the graph that we are going to consider in the rest of this paper. The non-physical phase space that we start with is therefore the space $\text{Cyl}_\Gamma$ of cylindrical functions on this graph $\Gamma$. This space is endowed with a Hilbert space structure inherited from the $\SU(2)$ Haar measure. As usual, $\text{Cyl}_\Gamma$ is isometric to the space
\be
\text{Cyl}_\Gamma\simeq\big(\text{Fun}(\SU(2))^{\otimes p},\de\mu_\Gamma\big),
\ee
where the measure $\de\mu_\Gamma\equiv\de\mu_0^{\otimes p}$ is defined as $p$ copies of the $\SU(2)$ Haar measure $d\mu_0$.

To construct the physical Hilbert space, we proceed as usual in two steps. The first one consists in implementing the Gauss constraint $G\approx0$ (\ref{gauss cons}), which leads to the gauge-invariant kinematical Hilbert space $\mathcal{H}^\text{kin}_\Gamma$. Any function $\psi\in\mathcal{H}^\text{kin}_\Gamma$ is invariant under the action of $\SU(2)$, generated by the Gauss constraint itself, at the unique vertex of the graph $\Gamma$. Therefore, we have the inclusion
\be\label{inclusion}
\mathcal{H}^\text{kin}_\Gamma\subset\big(\text{Fun}(\SU(2))^{\otimes p},\de\mu_\Gamma\big)/\SU(2),
\ee
where the quotient by $\SU(2)$ results from the invariance under the Gauss constraint. The measure $\de\mu_\Gamma$ is compatible with this quotient due to its right and left invariance properties.

So far, we have not specified the fact that the boundary is a black hole horizon, and this is the reason for which we have an inclusion in (\ref{inclusion}) and not a strict equality. In order to implement the black hole horizon condition, let us start by recalling the fact that the kinematical Hilbert space can be expanded in terms of spin network states. A spin network associated to the graph $\Gamma$ is an assignment of $\SU(2)$ unitary irreducible representations to the links of the graph, and an assignment of an $\SU(2)$ intertwiner to the vertex. In particular, the links $\ell$ crossing the horizon are colored by representations $j_\ell$. Due to the presence of the horizon, and because of the $\SU(2)$ invariance at the vertex, these representations $j_\ell$ are constrained to satisfy the two following requirements.
\begin{enumerate}
\item The first one is a consequence of the $\SU(2)$ gauge invariance, and we will refer to it as the invariance relation. Its significance is that the representations $j_\ell$ are such that there is an invariant tensor in the decomposition of their tensor product into irreducible representations. This can be written as
\be\label{invariantrelation}
\text{Inv}_{\SU(2)}\left(\bigotimes_{\ell=1}^pj_\ell\right)\neq\emptyset.
\ee
 \item The second one is a consequence of the finiteness of the length $L$ of the horizon, and we will refer to it as the length relation. Each link $\ell$ crossing the horizon carries a fundamental length $L_\ell=8\pi\lp\sqrt{j_\ell(j_\ell +1)}$ \cite{LFR} that contributes to the total length of the horizon according to
\be\label{lengthrelation}
L=8\pi\lp\sum_{\ell=1}^p\sqrt{j_\ell(j_\ell +1)}.
\ee
This relation is obviously the three-dimensional analogue of the one that fixes the macroscopic horizon area of four-dimensional black holes in loop quantum gravity.
\end{enumerate}
As a consequence, the kinematical Hilbert space $\mathcal{H}^\text{kin}_\Gamma$, in the presence of the black hole, is generated, as a vector space, by the spin networks such that the representations $j_\ell$ coloring the links $\ell$ crossing the horizon satisfy the invariance relation (\ref{invariantrelation}) and the length relation (\ref{lengthrelation}). The kinematical scalar product
trivially follows from the measure $\de\mu_\Gamma$ in (\ref{inclusion}).

At this point, one can wonder about the reason for which the length relation and the graph of figure \ref{graph} do indeed represent the tessellation of a black hole horizon, and not that of an arbitrary circle embedded into $\Sigma$. In other words, how can we be sure that we are going to compute the entropy associated with the horizon of a black hole, and not that of an arbitrary bounded region of perimeter $L$? The reason is that the number of microstates that we consider later on in (\ref{Nmicrostates}) corresponds to all possible recouplings (at a virtual intertwiner sitting at the center of the black hole) of the spin network links that cross the horizon. All these configurations have to be considered as different because what is inside the horizon is not accessible to an outside observer. If on the contrary we were describing a circle embedded into $\Sigma$, because of the topological nature of the theory a single graph with fixed coloring and fixed intertwiner would be sufficient in place of \ref{graph}.

Now that we have the kinematical setup, the next step is the construction of the physical Hilbert space $\mathcal{H}^\text{phys}_\Gamma$. This has to be done by implementing at the quantum level the remaining three constraints $H\approx0$ (\ref{curvature}). When the cosmological $\Lambda$ vanishes, we know how to impose these constraints and to construct explicitly the physical Hilbert space. In this case, the relation with the covariant quantization \`a la spin-foam is cristal-clear \cite{NP}, and the link with the combinatorial quantization \cite{MN1,MN2}, which is totally understood as well, exhibits a hidden quantum group symmetry. This quantum group, known as the Drinfeld double of $\SU(2)$ and usually denoted by $\DSU(2)$, is a deformation of the isometry group $\ISU(2)$ of the three-dimensional flat space $\mathbb{E}_3$. When the cosmological constant is non-vanishing, the resolution of the remaining constraints $H\approx0$ is much more involved than in the previous case, and, to our knowledge, it is still incomplete in the context of loop quantum gravity. However, there are strong indications about what the solutions of these constraints should look like in the case where $\Lambda=\ell_c^2$ is positive. Therefore, we will concentrate on this case in what follows. 

The imposition of the remaining three constraints $H\approx0$ has in fact two effects on the gauge-invariant kinematical Hilbert space. The first one concerns the invariance under spatial diffeomorphims. Indeed, it is well known that the constraints $H\approx0$ generate (on shell) space-time diffeomorphims, and that two out of the three components generate spatial diffeomorphisms. In fact, these two components do not depend on the cosmological constant. Their resolution is therefore independent of $\Lambda$, and implies that spin network states defined on homotopic graphs are identified at the physical level. This justifies a posteriori the use of a single (sufficiently refined) graph to capture all the physical degrees of freedom of the theory. Additionally, one can interpret the fact that these two components of $H\approx0$ do not depend on $\Lambda$ by the fact that, regardless of the value of the cosmological constant, the invariance under spatial diffeomorphims remains a symmetry of three-dimensional gravity in its Hamiltonian formulation.

The second effect arises when one imposes the third component of the constraints $H\approx0$. This is the only component in the set of constraints that depends explicitly on the cosmological constant. It is therefore natural to expect that the nature and the properties of its solutions will depend on the value of $\Lambda$. This is indeed the case, but unfortunately a precise resolution of this constraint in the context of loop quantum gravity is still missing. One can however guess the solution by looking at other quantization schemes of three-dimensional gravity. For instance, the path integral quantization \cite{Witten_Jones}, the Turaev-Viro model \cite{TV}, and the combinatorial quantization \cite{Schom,Schom2}, strongly suggest that in loop quantum gravity the physical states should be quantum spin networks, in the sense that they should be colored with $\text{U}_q(\su(2))$ representations on the links and $\text{U}_q(\su(2))$ intertwiners at the vertices of the graph $\Gamma$. The quantum group $\text{U}_q(\su(2))$ is defined for $q=\exp\big(i2\pi/(k+2)\big)$ a root of unity, and then the level $k=\ell_c/\lp$ is necessarily an integer. The Turaev-Viro model offers certainly the most concrete framework to see that physical states should be quantum spin networks, since these correspond precisely to the boundary states of the Turaev-Viro model. Nonetheless, it would be very interesting to see precisely how the resolution of the remaining constraint leads to the quantum group $\text{U}_q(\su(2))$. Some recent results \cite{Pranz1,Pranz2,Pranz3,Pranz4,Hanno} are first encouraging steps towards the resolution of this problem. For all these reasons, we choose the physical Hilbert space $\mathcal{H}^\text{phys}_\Gamma$ on the graph $\Gamma$ to be the one generated by the quantum spin networks. The spins $j_\ell$ are now interpreted as finite-dimensional representations of $\text{U}_q(\su(2))$, and therefore they cannot exceed $k/2$. This bound appears as a sort of infrared cut-off in the theory. In the same way, the unique vertex of $\Gamma$ is colored with a $\text{U}_q(\su(2))$ intertwiner, and the condition (\ref{invariantrelation}) has to be replaced, at the physical level, by the quantum invariant relation
\be\label{quantuminvariantrelation}
\text{Inv}_{\text{U}_q(\su(2))}\left(\bigotimes_{\ell=1}^pj_\ell\right)\neq\emptyset.
\ee
As a consequence, the physical Hilbert space is the subset
\be
\mathcal{H}^\text{phys}_\Gamma\subset\big(\text{Fun}(\SU_q(2))^{\otimes p},\de\mu_\Gamma^{(q)}\big)/\text{U}_q(\su(2)),
\ee
provided that the length condition (\ref{lengthrelation}) is satisfied when functions $f\in\mathcal{H}^\text{phys}_\Gamma$ are expanded into quantum spin networks. Here, $\text{Fun}(\SU_q(2))$ is the space of (polynomial) functions on $\text{U}_q(\su(2))$, and as a Hopf algebra is dual to $\text{U}_q(\su(2))$. The measure $\de\mu_\Gamma^{(q)}$ is a product of $p$ copies of the Haar measure $\de\mu_0^{(q)}$ on $\text{Fun}(\SU_q(2))$. The Haar measure $\de\mu_0^{(q)}$ is very similar to its classical counterpart $\de\mu_0$ in the sense that two quantum spin networks with different colors are orthogonal, as in the classical case. However, the norm of a quantum spin network state is different from the norm of the classical spin network state defined on the same colored graph. An explicit expression of the physical scalar product in not necessary for our purpose.

\section{The Bekenstein-Hawking entropy}
\label{sec:3}

\noindent Now we have all the ingredients necessary for the computation of the Bekenstein-Hawking entropy of the BTZ black hole from the point of view of loop quantum gravity. Indeed, when the cosmological constant is positive, the number $N_\Gamma(L,\Lambda)$ of states compatible with the relation (\ref{quantuminvariantrelation}) is given by
\be\label{Nmicrostates}
N_\Gamma(L,\Lambda)=\text{dim}\left(\text{Inv}_{\text{U}_q(\su(2))}\left(\bigotimes_{\ell=1}^pj_\ell\right)\right),
\ee
if the length relation (\ref{lengthrelation}) is satisfied, and it is null otherwise. When the length condition is satisfied, $N_\Gamma(L,\Lambda)$ can be written as the following finite sum:
\be\label{dimension}
N_\Gamma(L,\Lambda)=\frac{2}{k}\sum_{d=1}^{k+1}\sin^2\left(\frac{\pi}{k+2}d\right)
\prod_{\ell=1}^p\frac{\displaystyle\sin\left(\frac{\pi}{k+2}dd_\ell\right)}{\displaystyle\sin\left(\frac{\pi}{k+2}d\right)},
\ee
where $d_\ell=2j_\ell+1$ is the classical dimension of the spin-$j_\ell$ representation. Here $k$ is large and we will make the approximation $k+2\sim k+1\sim k$. As a consequence, the number of states becomes
\be\label{dimension approx}
N_\Gamma(L,\Lambda)\simeq\frac{2}{k}\sum_{d=1}^k\sin^2\left(\frac{\pi}{k}d\right)
\prod_{\ell=1}^p\frac{\displaystyle\sin\left(\frac{\pi}{k}dd_\ell\right)}{\displaystyle\sin\left(\frac{\pi}{k}d\right)}.
\ee
Before going to the analytic continuation, we would like to give an interpretation of this formula. To do so, let us stress that \eqref{dimension approx} needs two ingredients to be constructed:
\be\label{ingredients}
(i)\ \text{the sum}\qquad\frac{2}{k}\sum_{d=1}^k\sin^2\left(\frac{\pi d}{k}\right),\qquad
\text{and}\ (ii)\ \text{the characters}
\qquad\chi^{(j_\ell)}\left(\frac{\pi d}{k}\right)=\frac{\displaystyle\sin\left(\frac{\pi}{k}dd_\ell\right)}{\displaystyle\sin\left(\frac{\pi}{k}d\right)}.
\ee
The discrete sum $(i)$ is in fact the quantum analogue of the classical $\SU(2)$ Haar measure (restricted to gauge-invariant functions, it reduces to a one-dimensional integral over the $\SU(2)$ conjugacy classes), and it contains the information about the fact that we are considering the quantum group $\text{U}_q(\su(2))$ with $q$ a root of unity. One immediately sees that this quantum measure can be viewed as the Riemann sum approximating the classical $\SU(2)$ Haar measure. The characters $(ii)$ are also the quantum analogues of the classical $\SU(2)$ characters with discrete angles $\theta=\pi d/k$, and they contain the information about the type of representations that we are coloring the spin networks with. Here these representations are finite-dimensional.

Now the idea is to perform an analytic continuation of this formula to a negative value of $\Lambda$. In this case, the level $k$ becomes a purely imaginary integer, and we will denote it by $k=i\lambda$ where $\lambda >0$. To ensure that the analytic continuation is well-defined, one has to understand the upper bound of the sum in (\ref{dimension}) as the modulus $|k|$. This allows us to replace the upper bound $k$ by $\lambda$ in the sum, which then becomes restricted to the values $d\leq\lambda$. This is similar the restriction that has been used in the literature (see \cite{Carlip} for a review and also \cite{Chen}) to compute the number of black hole microstates from the point of view of the CFT. As a consequence, we define the number of the Euclidean BTZ black hole microstates by the formula
\be\label{N BTZ}
N^\text{BTZ}_\Gamma(L,\Lambda)\simeq\frac{2}{\lambda}\sum_{d=1}^\lambda\sinh^2\left(\frac{\pi}{\lambda}d\right)
\prod_{\ell=1}^p\frac{\displaystyle\sinh\left(\frac{\pi}{\lambda}dd_\ell\right)}{\displaystyle\sinh\left(\frac{\pi}{\lambda}d\right)},
\ee
where $\Lambda=-\ell_c^2$ is now negative. Note that we have also replaced the factor $1/k$ in the measure (\ref{ingredients}) by $1/|k|$, which becomes $1/\lambda$ after the analytic continuation. By doing this, $N^\text{BTZ}_\Gamma(L,\Lambda)$ remains an integer, and it can be interpreted as a number of states. This is exactly the expression that has been introduced in \cite{US} to compute the entropy of four-dimensional black holes in loop quantum gravity with a complex Barbero-Immirzi parameter.

In the classical limit, i.e. when the representations $j_\ell$ become large and $\lp$ approaches zero with the product $\lp j_\ell$ remaining finite, the sum (\ref{N BTZ}) is dominated by the term $d=\lambda$. Therefore, we have
\be
N^\text{BTZ}_\Gamma(L,\Lambda)\simeq\frac{2}{\lambda}\sinh^2(\pi)
\prod_{\ell=1}^p\frac{\sinh({\pi}d_\ell)}{\sinh(\pi)}\simeq\frac{2}{\lambda}\sinh^2(\pi) 
\prod_{\ell=1}^p\exp\big(\pi (d_\ell-1)\big).
\ee
As a consequence, it is immediate to show that the entropy $S_\Gamma^\text{BTZ}(L,\Lambda)=\log\left(N^{BTZ}_\Gamma(L,\Lambda)\right)$ is equivalent in the classical limit to
\be
S_\Gamma^\text{BTZ}(L,\Lambda)\sim\frac{L}{4\lp},
\ee
where we have used the length relation (\ref{lengthrelation}). We recover exactly the Bekenstein-Hawking formula for the entropy.

\section*{Discussion}

\noindent In this work, we have proposed a description of the BTZ black hole in the context of loop quantum gravity. As in the four-dimensional case, the black hole horizon is viewed as a (circular) boundary on the spacelike surface $\Sigma$, and kinematical states are cylindrical functions on a graph embedded in $\Sigma$. We have argued that, due to the topological nature of three-dimensional gravity (in the bulk and not on the horizon), a graph $\Gamma$ that consists in an arbitrary number $p$ of links crossing the horizon and meeting at the spatial infinity at one vertex only, is sufficient to capture all the degrees of freedom of the theory. Then, we have given a precise definition of the kinematical Hilbert space $\mathcal{H}^\text{kin}_\Gamma$, and constructed, when the cosmological constant $\Lambda$ is positive, the physical Hilbert space $\mathcal{H}^\text{phys}_\Gamma$  in terms of the representation theory of $\text{U}_q(\su(2))$ with $q$ a root of unity. For a fixed configuration $(j_1,\dots,j_p)$ of spins coloring the links at the horizon, we have computed the number $N_\Gamma(L,\Lambda)$ of physical states, and then suggested an analytic continuation of this formula to negative values of $\Lambda$. The resulting function $N^\text{BTZ}_\Gamma(L,\Lambda)$ has been interpreted as the number of BTZ black hole microstates associated to the given configuration, and we have finally shown that, in the classical limit, the leading order term in $S=\log\left(N^\text{BTZ}_\Gamma(L,\Lambda)\right)$ reproduces exactly the Bekenstein-Hawking for the entropy for all values of $p$. To our knowledge, it is the first time that a model is proposed for the computation of the BTZ black hole entropy in the context of loop quantum gravity.

Even if our derivation relies on certain assumptions, this striking result strongly suggests that it might be possible to recover the entropy of the BTZ black hole from three-dimensional loop quantum gravity. Furthermore, it is very interesting to notice that the techniques used here in three dimensions are very similar to the ones that were used to compute the entropy of four-dimensional black holes \cite{US}. In this sense, there is a kind of universality that certainly deserves to be explored in more details. Our proposal raises many questions that we should investigate more carefully. We now list a few of them, together with tentative answers.

\subsubsection*{Why are we indeed computing the partition function of the BTZ black hole?}

\noindent This first question concerns the physical meaning of the analytic continuation that we have introduced. To give a physical interpretation, let us recall that three-dimensional gravity is equivalent to a Chern-Simons theory whose gauge group is exactly the isometry group of the local solutions to the Einstein equations. This isometry group depends obviously on the sign of the cosmological constant $\Lambda$, and also on the signature $\sigma\in\{\text{E},\text{L}\}$ ($\text{E}$ and $\text{L}$ stand for Euclidean and Lorentzian, respectively) of the space-time. As it can be clearly seen from the combinatorial quantization scheme (see \cite{Schom,Schom2} for instance), the quantization of Chern-Simons theory turns these classical groups into quantum groups according to table \ref{table}.
\begin{table}
\begin{tabular}{|c|c|c|}
\hline
& $\sigma$ = E & $\sigma$ = L \\
\hline
~ $\Lambda=0$ ~ & $\DSU(2)$ & $\DSU(1,1)$ \\
\hline
$\Lambda>0$ & $\quad$ $\text{U}_q(\so(4))$, $q$ root of unity $\quad$ & $\text{U}_q(\so(3,1))$, $q$ real \\
\hline
$\Lambda<0$ & $\text{U}_q(\so(3,1))$, $q$ real & $\quad$ $\text{U}_q(\so(2,2))$, $q$ a phase $(|q|=1)$ $\quad$ \\
\hline
\end{tabular}
\caption{Quantum groups of three-dimensional quantum gravity for different signs of the cosmological constant and Euclidean or Lorentzian signature.}\label{table}
\end{table}
In this table, $\text{U}(\mathfrak{g})$ denotes the enveloping algebra of the Lie algebra $\mathfrak{g}$, $\DSU$ denotes the quantum double of the enveloping algebra, and $\text{U}_q(\mathfrak{g})$ denotes the quantum deformation of the classical enveloping algebra. The analytic continuation that we have introduced in this paper has two ingredients that seem to fit naturally in the previous table. First, we make the level purely imaginary, and this maps Euclidean quantum gravity with $\Lambda >0$ to Euclidean quantum gravity with $\Lambda<0$. However, it is known from the combinatorial quantization scheme \cite{BNR} that the quantization of Euclidean gravity with negative $\Lambda$ leads to an infinite-dimensional Hilbert space. More precisely, as $q$ is real in this case, there is no upper cut-off for the representations, and the formal expression for the dimension of the Chern-Simons Hilbert space diverges. This brings us to the second ingredient: while we make the level purely imaginary we keep the cut-off $|k|$ in the sum  defining the quantum measure (\ref{ingredients}). Things  happen as if we had turned again $q$ to a root of unity: this takes us from the Euclidean regime to the Lorentzian one, where the BTZ black hole makes sense as a classical solution. Even if in the Lorentzian regime with $\Lambda<0$ the deformation parameter $q$ is generically a phase, many arguments have arrived at the conclusion that in the presence of a BTZ black hole $q$ should be a root of unity  \cite{Chen}. All this suggests that one could interpret our construction as a recipe that sends the partition function of Euclidean gravity with $\Lambda>0$ to that of Lorentzian gravity with $\Lambda<0$ in the presence of a BTZ black hole. This can be summarized by the following formal diagram:
\be
\mathcal{Z}_\text{E}^{(+)}=\int\mathcal{D}g\,\exp\left(iS_\text{E}^{(+)}[g]\right)\longrightarrow
\mathcal{Z}_\text{E}^{(-)}=\int\mathcal{D}g\,\exp\left(iS_\text{E}^{(-)}[g]\right)\longrightarrow
\mathcal{Z}_\text{L}^{(-)}=\int\mathcal{D}g\,\exp\left(iS_\text{L}^{(-)}[g]\right), 
\ee
where $S_\sigma^{(\pm)}[g]$ denotes the action for gravity with signature $\sigma$ and cosmological constant $\Lambda$ such that $\text{sign}(\Lambda)=\pm$,
and $\mathcal{Z}_{\sigma}^{(\pm)}$ is the associated partition function. The first arrow corresponds to turning $q$ root of unity to $q$ real, and the second arrow corresponds to keeping the bound in the sum (\ref{ingredients}) defining the quantum Haar measure. Of course, for the moment this is just an interpretation that certainly deserves to be investigated deeper. However, it is interesting to note that the same kind of arguments appear in the CFT approach. We hope that our new way of deriving the entropy of the BTZ black hole will open a new path towards a complete description of the quantum microstates for the black hole. 

\paragraph{Why do we only need one graph to compute the entropy?}

\noindent This second question concerns the eventual computation of the total number of the black hole microstates. Indeed, in this paper, we have computed only the number of microstates for a given configuration $\Gamma$. If we mimic the techniques used for computing the entropy of four-dimensional black holes in loop quantum gravity, we should define the total number $N(L,\Lambda)$ of microstates as the sum
\be
N(L,\Lambda)=\sum_\Gamma w_\Gamma N_\Gamma(L,\Lambda),
\ee
where $w_\Gamma$ is a weight associated to the configuration $\Gamma$. The sum over $\Gamma$ means that we sum a priori over the number of links $p$, and also over the representations $(j_1,\dots,j_p)$ coloring these links. However the situation is much simpler in three dimensions because of the topological nature of the theory. Indeed, one expects that the Hilbert spaces $\mathcal{H}^\text{phys}_\Gamma$ are all physically equivalent for different graphs $\Gamma$, as long as they have enough structure to capture the relevant topological data of the spacelike surface $\Sigma$. One strong indication of this equivalence lies in fact in the main result of this paper, which is that the number $N^\text{BTZ}_\Gamma(L,\Lambda)$ of microstates is indeed independent of $\Gamma$ in the large $L$ limit. 
One therefore expects to be able to describe  the physical Hilbert space for the BTZ black hole  from the analytic continuation of a single physical Hilbert space $\mathcal{H}^\text{phys}_{\Gamma}$ for a single graph $\Gamma$. 

\paragraph{Can we establish a contact with the CFT approach?}

\noindent If we follow the logic of the previous paragraph, then it is natural to consider the simplest possible graph $\Gamma$ to construct the physical Hilbert space. It is clear that the simplest graph consists in only two links crossing the horizon, and meeting at one vertex at infinity. When the cosmological constant $\Lambda$ is positive, the graph with only one link is necessarily trivial in the sense that the representation labeling the only link must be trivial. When the graph contains two links, they should be colored by the same representation (at least when there is no angular momentum). We will come back to this observation a bit later. For the moment, we assume that the graph $\Gamma$ that consists of two links is sufficient to describe the physical states, and the two links can be colored by any pair $(j_1,j_2)$ of representations. Let us denote by $\mathcal{H}^\text{phys}(j_1,j_2)$ the associated Hilbert space, and by $N(j_1,j_2)$ its dimension. Now several questions naturally arise. What is the physical meaning of the representations $j_1$ and $j_2$? Are all the states in $\mathcal{H}^\text{phys}(j_1,j_2)$ physically inequivalent? Or, in the language of CFT, does $\mathcal{H}^\text{phys}(j_1,j_2)$ contain zero norm states?

To answer these questions, it would be worth trying to establish a contact between our loop quantum gravity description of the BTZ black hole and the standard CFT approach. In the later, the computation of the asymptotic number of black hole microstates relies on the Cardy formula for the density of states in a two dimensional CFT \cite{Cardy}. In summary (see \cite{Carlip} for instance), the total number of black hole microstates is given by the number $\rho(\Delta,\overline{\Delta})$ of CFT states, where $\Delta$ and $\overline{\Delta}$ are eigenvalues of the standard Virasoro operators $L_0$ and $\overline{L}_0$, and
\be
\Delta=\frac{\ell_cM+J}{2},
\qquad\qquad\text{and}\qquad\qquad
\overline{\Delta}=\frac{\ell_cM-J}{2}.
\ee
$\Delta$ and $\overline{\Delta}$ are totally fixed by the value of the cosmological constant $\Lambda=-\ell_c^2$, the mass $M$, and the angular momentum $J$ of the black hole. Using the Cardy formula \cite{Strom} and its subleading corrections \cite{Carlip_log}, one can show that
\be
\log\big(\rho(\Delta,\overline{\Delta})\big)\sim\frac{L}{4\lp}-\frac{3}{2}\log\left(\frac{L}{\lp}\right). 
\ee
In this way, one recovers the Bekenstein-Hawking entropy and its logarithmic corrections. Note that the corrections were also obtained in \cite{logcorr}. As a consequence, it seems that our number of states $N(j_1,j_2)$ plays a role similar to the density of states $\rho(\Delta,\overline{\Delta})$, in the sense that these two quantities depend on two parameters and have the same leading order term in the classical limit. We can go even further to stress this similarity. Indeed, when there is no angular momentum, i.e. $J=0$, we have argued above that the representations $j_1$ and $j_2$ should be equal to the same value, let us say $j_1$, which is related in the classical limit to the length $L$ of the horizon by the relation $16\pi\lp j_1=L$, and $j_1=r_+/(8\lp)$. From the CFT point of view, it also appears that $\Delta=\overline{\Delta}$ when the angular momentum vanishes, and $\Delta=3r_+^2/(32 \lp^2)$. This is consistent with the idea that the representations $j_1$ and $j_2$ could play the same role as $\Delta$ and $\overline{\Delta}$, and should be fixed by the values of the mass $M$ and the angular momentum $J$ of the BTZ black hole. From the non-rotating case, we see that $j_1\propto\sqrt{\Delta}$. What happens when the angular momentum is not vanishing? It has been suggested recently \cite{AleErnesto} that in the loop quantum gravity description of four-dimensional black holes, an angular momentum can be taken into account by the presence of an extra puncture (crossing the horizon) whose representation is proportional to the value of the angular momentum $J$. If we assume that the same phenomenon exits in our three-dimensional model, then the representation $j$ coloring the extra-puncture should satisfy the inequality $j\leq j_1+j_2\propto\sqrt{M}$, which is to be contrasted with the inequality $|J|\leq M\ell_c$ arising from classical gravity.

\paragraph{Logarithmic corrections and large diffeomorphisms.}

\noindent Now, let us comment on the question concerning the equivalence or non-equivalence between states in $\mathcal{H}^\text{phys}(j_1,j_2)$. This question comes naturally when one notices that the dimension $N(j_1,j_2)$ of $\mathcal{H}^\text{phys}(j_1,j_2)$ is larger than the number of states $\rho(\Delta,\overline{\Delta})$, since
\be
\exp\left(-\frac{L}{4\lp}\right)\left(N(j_1,j_2)-\rho(\Delta,\overline{\Delta})\right)\sim\frac{3}{2}\log\left(\frac{L}{\lp}\right)
\ee
due to the logarithmic corrections arising in the CFT approach. This means that we are counting more states in our model. This discrepancy could originate from a forgotten symmetry in loop quantum gravity. Indeed, it has been argued in the CFT approach that the logarithmic corrections are intimately related to the modular invariance of the theory, and the modular invariance represents the invariance under large diffeomorphims. These large symmetries have never been taken into account in our approach, and they could be easily considered. Indeed, they might be related to the natural braided statistics associated to the quantum groups we are dealing with. We could expect the action of the braiding group in our approach to be as important as the modular invariance in the CFT approach. The introduction of a non-trivial braiding should  not change the leading order behavior of the entropy in the classical limit (because the braiding group is discrete). However, it might be important (even crucial) for the recovery of the expected logarithmic corrections to the entropy.

\paragraph{What are the physical microstates of the black hole?}

\noindent  An important question remains concerning the  physical interpretation of these microstates. How is it that a topological quantum field theory can provide such a large number of states for the black hole, and what is the relation with the underlying microscopic physical degrees of freedom of the BTZ black hole? These questions  remains open so far. From the mathematical point of view, one understands that this huge number of states could come from the quantization of the CFT at the boundary. But, from the physical point of view, it is not totally satisfactory to think that the black hole microstates come from a theory that lives very far from the horizon. 

In our approach, the degrees of freedom seem to live on the black hole boundary, but we still do not know what they are. It might be that these states result from all the matter fields that contributed to forming the black hole during its gravitational collapse, and which have disappeared behind the horizon. To verify this hypothesis, one should study precisely the gravitational collapse of the BTZ black hole from the point of view of loop quantum gravity. It is important to keep in mind that such questions are also open in the four-dimensional models. The encouraging point is that a definite answer might very well be within reach in the simplified context of three-dimensional gravity, and this could provide valuable guiding insights for the physical four-dimensional theory. We leave these investigations for future work.

\subsection*{Aknowledgments}

\noindent We would like to thank the loop quantum gravity team in Marseille, and especially Carlo Rovelli and Simone Speziale for the enthusiasm they have shown when this work was presented. We also thank Marc Henneaux for pointing out reference \cite{BHTZ}. MG is supported by the NSF Grant PHY-1205388 and the Eberly research funds of The Pennsylvania State University.

\end{document}